\documentclass[twoside,a4paper,10pt]{amsart}

\usepackage{amsmath}
\usepackage{amssymb}
\usepackage{amsthm}
\usepackage{mathrsfs}

\newcommand{\comment}[1]{}

\newcommand{\pd}[2]{\frac{\partial {#1}}{\partial {#2}}}
\newcommand{\n}{\noindent}
\newcommand{\sgn}{\textrm{sgn}}
\newcommand{\RE}{\mathbb R}
\newcommand{\CO}{\mathbb C}
\newcommand{\ve}{\varepsilon}

\renewcommand{\Im}{\operatorname{Im}\,}

\newcommand{\XX}{\mathcal X}

\newcommand{\VV}{\overline V}

\newcommand{\oV}{\overline V}

\newcommand{\Hboh}{H^{\oV}}
\newcommand{\Rboh}{R^{\oV,\ve}}

\newtheorem{theorem}{Theorem}
\newtheorem{lemma}{Lemma}
\newtheorem{proposition}{Proposition}

\theoremstyle{definition}
\newtheorem{example}{Example}

\newcommand{\erre}{\mathbb{R}}

\newcommand{\de}{\delta} \newcommand{\al}{\alpha}
\newcommand{\la}{\lambda}

\newcommand{\ba}{\begin{eqnarray}} \newcommand{\ea}{\end{eqnarray}}
\newcommand{\bdm}{\begin{displaymath}}
\newcommand{\edm}{\end{displaymath}} \newcommand{\brr}{\begin{array}}
\newcommand{\err}{\end{array}} 
 \newcommand{\lf}{\left}
\newcommand{\ri}{\right}

\usepackage{mathrsfs}

\newcommand{\bml}{\begin{gather}} \newcommand{\eml}{\end{gather}}

\newcommand{\ga}{\gamma}
\newcommand{\Ga}{\Gamma}

\newcommand{\oH}{\overline H}
\newcommand{\ulim}{\operatorname{u}-\lim}

\newcommand{\gsmt}{\tilde\gamma}
\newcommand{\Osmt}{\tilde\Omega}
\newcommand{\Hsmt}{\tilde H}
\newcommand{\Vsmt}{\tilde V}
\newcommand{\Rsmt}{\tilde R}
\newcommand{\Gsmt}{\tilde\Gamma}
\newcommand{\Ismt}{\tilde I}
\newcommand{\bsmt}{\tilde b}

\begin{document}

\title[ ]{Coupling in the singular limit of thin quantum waveguides}

\author{Sergio Albeverio}
\address{Albeverio: Institut f\"ur Angewandte Mathematik}
\curraddr{Wegelerstr. 6, 53115 Bonn, Germany}

\email{albeverio@wiener.iam.uni-bonn.de}

\author{Claudio Cacciapuoti} 
\address{Cacciapuoti: Institut f\"ur Angewandte Mathematik}
\curraddr{Wegelerstr. 6, 53115 Bonn, Germany}

\email{caccia@na.infn.it}

\author{Domenico Finco} 
\address{Finco: Institut f\"ur Angewandte Mathematik}
\curraddr{Wegelerstr. 6, 53115 Bonn, Germany}

\email{finco@wiener.iam.uni-bonn.de}



\thanks{\n The authors thank Pavel Exner and Rodolfo Figari for useful comments and discussions. \newline
\n This work was 
supported by the 
EU-Project ``Quantum Probability with Applications to Physics,
Information Theory and Biology'', by the Collaborative Research Center
(SFB) 611 ``Singular Phenomena and Scaling in Mathematical Models''
and by the Deutsche Forschungsgemeinschaft (DFG)}

{\maketitle}

\begin{abstract}
We analyze the problem of approximating a smooth quantum waveguide with
a  quantum graph. We consider a planar curve with
compactly supported curvature and a 
strip of constant 
width around the curve. We rescale the curvature and the width in such a 
way that
the strip 
can be approximated by a singular limit curve, consisting of one
vertex and two infinite, straight edges, i.e. a broken line. We discuss
the convergence of the Laplacian, with Dirichlet boundary conditions on
the strip, in a suitable sense
and we obtain two possible
limits: the Laplacian on the line with Dirichlet boundary conditions in
the origin and a non trivial
family of point perturbations of the Laplacian on the
line. The first
case generically occurs and
corresponds to the decoupling of the two components of the limit
curve, while in the second case a coupling takes place.  We present also
two families of curves which give rise to coupling. 
\end{abstract}

\section{Introduction}

\n In many microscopic systems a quantum
particle is constrained by a 
 confining potential to a region
with transversal dimensions small with respect to the longitudinal
ones.  For example in organic molecules the atoms make strong bonds
and  organize  themselves
on a regular structure, then the
$\pi$-electrons 
move in correspondence of the bonds under the action of a strong confining
potential. Since the early 50s one dimensional models were used to
describe the dynamics of $\pi$-electrons in such molecules (see,
e.g., \cite{RS}).  

\n In more recent times a growing interest in the quantum dynamics of
particles in quasi one 
dimensional structures has been driven by the possibility to realize
devices with transversal
dimensions on the scale of length of hundreds of nanometers, such as
nanotubes or quantum wires.  The
possibility of a ``nanotechnology'' was already  
envisaged by R.
Feynman in  1959
(see \cite{Fy}), but the turning point can be fixed 
 in 1981 when  G. Binnig and H. Rohrer, of IBM's Z\"urich
Lab,  invented the 
scanning tunneling microscope, making it possible to inspect and manipulate
matter on the atomic scale. 

\n Quantum-graphs represent an excellent 
model for many quasi one dimensional structures  
like organic molecules, nanotubes and systems of quantum wires.  In
mathematical terms  a 
quantum-graph is realized by a graph (i.e. a  
set of points, the vertices, and a set of finite or infinite segments that connect the vertices, the edges),
together with a quantum
dynamics for a particle on the graph generated by  self-adjoint
differential or
pseudo-differential operators on the graph (see \cite{K02}, \cite{K04} and \cite{K05}).  

\n From the  point of view of
mathematical physics it is an open question to understand in which
sense the one dimensional dynamics on a  quantum graph approximates the
dynamics of a particle 
constrained  on a region with 
small transversal dimensions. Essentially one can isolate two
problems: to determine which 
one dimensional, differential or pseudo-differential
operators are most suitable in order to describe the
dynamics on the edges and 
which couplings in the vertices among the edges can  be physically
feasible.

\n A strategy to approach both these open  problems, in the case of
differential operators, consists in
studying the limit of the operator minus the Laplacian 
defined on two or three dimensional domains with a graph-like
topology but finite width, when the width
goes to zero.

\n This paper deals with the problem of the coupling in the
vertices. For this reason we want to consider the simplest possible
limit dynamics on the edges. We take a planar domain of constant
width and  which is straight
outside a  compact region. It is known
that  for such  a  kind of domain the limit dynamics on the straight part of the edges will be
generated  by the one dimensional Laplacian (see e.g. 
\cite{KZ}, \cite{RuSc},
\cite{Sa}, \cite{EP}, \cite{Po06} for the Neumann case, \cite{Po} for the
Dirichlet case with a narrowing producing decoupling,  and 
\cite{DT} for the  case with quadratic confining potentials). 

\n In the case of a graph with the free Laplacian on the edges there exists a
complete characterization of all the possible 
couplings in the vertices (see \cite{KS}). To define 
the coupling in 
a vertex there are   
$n^2$ real parameters at disposal, where $n$ is the number of
the edges connected with  the vertex. It is not clear which boundary
conditions 
can be obtained as the result of taking the zero width limit from a strip
or a cylinder around the graph and how the 
parameters are related to physical properties of the system such as the
geometry of the graph 
(see, e.g.,
\cite{ES} and the appendix by P. Exner in \cite{AGH-KHII}).

\n The 
problem of the convergence in the vertices strongly depends   on the
conditions imposed on the
boundary of the domain with finite thickness. Some well established results
exist  in the case with Neumann boundary conditions (see,
e.g., \cite{Bon},  \cite{EP}, \cite{KZ},  \cite{Po06}, \cite{RuSc},
 \cite{Sa}). All the results
 indicate that the coupling in 
the vertices is of Kirchhoff type, i.e., the  wave function is continuous in
the vertices and the  sum
over all the  first derivatives
of the wave functions on the edges connected to a vertex is equal
to zero. 

\n Analogous results do not exist for the case with Dirichlet
boundary conditions. This case is
discussed, for vertices with any number of edges, in the work by
O. Post \cite{Po}.  There the problem of a manifold
shrinking to a  graph is analyzed.
Under the hypothesis that the manifold 
narrows around the vertices, it is proved that the
spectrum of the operator minus the Laplacian on the manifold
 converges to the spectrum
of minus the Laplacian on the graph with decoupling boundary
conditions in the vertices, i.e. wave function equal to zero in the
vertices. See also the work by S. Molchanov and B. Vainberg \cite{MVpp} 
for the analysis of the scattering problem in the Dirichlet case.

\n The case with Dirichlet boundary conditions is physically very 
relevant and of great interest, because it correctly describes 
particles confined in a bounded region. It is reasonable  to believe that,
also in this case, it
is possible to obtain non decoupling conditions in the vertices.  

\n  The
difficulties arising in the Dirichlet case, with respect 
to the Neumann one, are related to the spectrum of the
Laplacian on compact domains. Only in the Neumann case, zero
is an  eigenvalue and the constant function is the corresponding
eigenfunction.  The occurrence of  the
eigenvalue zero makes 
it possible to 
approximate the wave function by a constant in a
small neighborhood of the vertices, that is crucial to prove the
convergence to Kirchhoff type 
conditions in the vertices.
Such a simple approximation does not hold in the Dirichlet case and at
the present time a reasonable guess on how to
approximate the wave
function near the vertices is lacking. 
 
\n As a first step in the analysis of the Dirichlet case we consider
a simple case of a planar quantum 
waveguide, i.e. we consider a strip in the plane with constant width 
 around a smooth curve and we take the Laplacian with Dirichlet
boundary conditions on this domain. In such a case it is possible to
define a system of 
global coordinates given by the arc length of the curve and the
distance from the curve (such a natural system of global coordinates
does not exist for a general domain). 

\n In our model the  quantum
waveguide will 
``collapse''  on a ``prototypical'' quantum graph made up of a
broken line, this is achieved with a suitable scaling 
of the width and of the curvature of the strip. We
assume that the curve is 
a straight line outside of a compact region, i.e. the signed curvature
$\gamma(t)$, $t\in\erre$, is a function with compact support. We
introduce a dimensionless 
scaling parameter, $\ve$, and assume that the width of the waveguide
scales as $\ve^\alpha d$, where $d$  is a positive 
constant and $\alpha\geqslant1$, while  the curvature scales as
$\ve^{-1}\gamma(t/\ve)$. Under these 
assumptions, when $\ve$ goes to zero, the waveguide narrows to a one
dimensional domain made up of two straight lines with the same
origin. Let us notice that with this scaling the 
angle between the straight parts of the curve is fixed.

\n Our main result, stated in theorem \ref{mainth}, is the following:
for $\alpha>5/2$ generically the limit  operator corresponds to the
free Laplacian with 
decoupling boundary conditions in the origin; nevertheless,  
if the curvature is such that the one dimensional Hamiltonian
$-\Delta-\gamma^2/4$ has a zero energy resonance, the limit operator
is a point perturbation of the Laplacian in dimension one
and the boundary conditions are non decoupling. 

\n We prove
the uniform convergence of the resolvent. A 
renormalization of the spectral parameter is necessary because of the
divergence of the term corresponding to the kinetic energy associated
with the motion in the transversal direction relative to the
  curve. This 
renormalization procedure has been used before when dealing  with Dirichlet
boundary conditions or with quadratic confining potentials (see
\cite{DT} and \cite{Po}).

\n We consider two examples of curves that generate a non decoupling
dynamics. Such examples indicate that the angle $\theta$
between the straight parts of the curve,
is not enough to characterize the limit dynamics. An interesting open
question is to understand which geometrical quantities are sufficient
to characterize the limit dynamics. 

\n For a simple case of a quantum graph we obtained, for the first time, non
decoupling boundary conditions in the vertex in the Dirichlet
case. The uniform
convergence of the resolvent can be a first step to prove
the convergence of the dynamics. 

\n Our result is consistent
with the one obtained by O. Post (see 
\cite{Po}). In fact in that work the decoupling in the vertices was
due to the narrowing of the domain 
in a neighborhood  of the vertices.

\n Our model is basically the same as in \cite{DT}. In their paper
G. Dell'Antonio and L. Tenuta studied the case in which the particle is
confined around a curve by a quadratic 
potential and they focused their attention on
the convergence of the 
dynamics on the strip to the dynamics on the broken line. As intermediate
step they proved that the
quadratic confining potential is 
 equivalent to a domain with Dirichlet boundary condition. Moreover
 they proved that in the general case the limit dynamics is decoupling. 

\n It is not trivial to extend our
result to the case of three or more edges connected to the same
 vertex, a result weaker than the convergence of the resolvent (e.g. the 
convergence of the spectrum) would already be of great interest. Nevertheless
we think that, similarly as in the case we discuss here, the role played by the
resonances will be   
decisive, even for proving a weaker result.

\n The paper is organized as follows. In section \ref{sec1}  we describe
in detail our model and we state the main  theorem. The section
\ref{sec2} is devoted to the proof of the main theorem. In section
\ref{sec3}  we give a characterization of the limit operator: we
discuss its spectrum, we give the integral kernel
of the propagator and evaluate the elements of the scattering
matrix. In the last section we investigate the relation between the
curvature and the limit dynamics and describe two  examples of curves
that generate a non decoupling limit dynamics.

\section{\label{sec1}Main theorem}

\n
In this section we shall present our main theorem. First we shall
introduce our model 
of a quantum strip and we shall recall some basic facts about the low
energy expansion of  the resolvent of a one dimensional Schr\"odinger operator.

\n
Let $\Ga$ be a curve in $\erre^2$ given in parametric form by $\Ga= \{ ( \ga_1(t), \ga_2(t) ), t\in \erre \}$
and let us assume that it is parameterized by the arc length $t$, i.e. $ \ga'_1(t)^2+ \ga'_2(t)^2=1$. 
We also introduce the signed curvature
\begin{equation*}
\gamma(t)=\gamma'_{2}(t)\gamma''_{1}(t)
-\gamma'_{1}(t)\gamma''_{2}(t)\,;
\end{equation*}
\n the curvature radius of $\Ga$ in $t$
 is equal to the inverse of the modulus of the signed curvature.

\n
We shall assume that $\ga$ has compact support, therefore $\Ga$ is a straight line
outside a compact region. We shall assume also that $\Ga$ has no self-intersection. Thus
$\Ga$ consists of two straight lines, $l_1$ and $l_2$, with the
origins, $O_1$ and $O_2$, connected by a piecewise $C^4$, non self-intersecting, curve $C$, running in a compact region. The integral of
$\ga$ gives the angle $\theta  $  
between  $l_1$ and $l_2$.

\n
Let us denote by $\Omega$ the strip of width $2d>0$ around $\Ga$:
\begin{equation*}
\Omega= \{ (x,y)\,\,s.t.\,\,  x= \ga_1(t) -  s \ga_2'(t), y= \ga_2(t) +  s \ga_1'(t), t\in\erre, s\in[-d,d] \}\,.
\end{equation*}
We assume $\sup_t |\ga(t)|d < 1$, in this way $(t,s)$ provide a global
system of coordinates in $\Omega$.

\n
We denote the Laplacian with Dirichlet boundary condition on
$\partial\Omega$ by $-\Delta^D_{\Omega}$;  
$-\Delta^D_{\Omega}$ is defined as the Friedrichs extension of $-\Delta$ 
with domain $C^{\infty}_0(\Omega \setminus \partial \Omega)$.

\n
It is convenient to write $-\Delta^D_{\Omega}$ in terms of  the
curvilinear coordinates $(t,s)$.  In particular the following
proposition holds (see \cite{DE} and \cite{ESjmp} for more details)

\begin{proposition}\label{prop1}
Assume that $\Ga$ has no self-intersections,
let $\ga $ be piecewise $C^2$ with compact support and $\ga',\ga''$ be bounded, then $-\Delta^D_{\Omega}$ is unitarily equivalent to 
the operator $H$ which is defined as the closure 
of the essentially self-adjoint operator 
$H_0$ acting on $L^2(\erre \times [-d,d])$ defined by
\begin{equation*}
H_{0}=
-\pd{}{t}\frac{1}{(1+s\gamma(t))^2}\pd{}{t}-
\pd{^2}{s^2}+V(t,s)\,,\qquad t\in\erre,\;s\in[-d,d]\,,
\end{equation*}
\n with
\begin{equation*}
V(t,s)=-\frac{\gamma(t)^2}{4(1+s\gamma(t))^2}
+\frac{s\gamma''(t)}{2(1+s\gamma(t))^3}
-\frac{5}{4}\frac{s^2\gamma'(t)^2}{(1+s\gamma(t))^4}
\end{equation*}
\n and domain given by
\begin{equation*}
{\mathscr D}(H_{0})=\Big\{ \psi\in L^2(\RE\times[-d,d]) \,\,
s.t. \,\, \psi\in C^{\infty } (\RE\times[-d,d])\,,\, \psi(t,d)=\psi(t,-d)=0 \, , \,
H_0 \psi \in L^2(\RE\times[-d,d]) 
\Big\}\,.
\end{equation*}
\end{proposition}
\n With piecewise $C^2$ we mean a function which is continuous, with continuous first derivative and eventually with a finite number of discontinuities in the second derivative.

\n
Let us also introduce the normal modes, 
that is the orthonormal complete system $\{\phi_n\}$ in $L^2([-d,d])$ whose elements
satisfy the following equation:
\begin{equation*}
\left\{\begin{aligned}
&-\frac{d^2}{ds^2}\phi_{n}=
\la_n \phi_{n}\\
&\phi_{n}(-d)=\phi_{n}(d)=0
\end{aligned}\right.\qquad 
n=1,2,3,\dots\,.
\end{equation*}
It is straightforward to compute $\phi_{n}$ and $\la_n$ explicitly
\begin{equation*}
\la_n = \Big(\frac{n\pi}{2d}\Big)^2 \qquad \phi_{n}(s)= 
\left\{
\begin{aligned}
\frac{1}{d^{1/2}} \cos \lf( \frac{n\pi s}{2d} \ri)\,, &\qquad  n\,\,
\text{odd} \\ 
\frac{1}{d^{1/2}} \sin \lf( \frac{n\pi s}{2d} \ri)\,, & \qquad n\,\,
\text{even}\,.
\end{aligned}
\ri.
\end{equation*}
\n
We rescale $\ga$ and $d$ in the following way:
\begin{equation*}
\left\{\begin{aligned}
&\ga(t) &\longrightarrow &\,\,\frac{1}{\ve} \ga\lf( \frac{t}{\ve} \ri) \\
&d &\longrightarrow &\,\,\ve^{\al} d
\end{aligned}\right.\qquad\ve>0\,,\;\alpha\geqslant1\,.
\end{equation*}
In this way we obtain a family of domains $\Omega_{\ve}$ and of operators $-\Delta^D_{\Omega_{\ve}}$
such that $\Omega_{\ve}$ approximates, for $\ve \rightarrow 0$, the
broken line of angle $\theta$ made up by $l_1$ and
$l_2$ with the same origin, $O_1\equiv O_2\equiv O$. 
Notice that the angle $\theta$ is unchanged by  the rescaling. We assume $\al \geqslant 1$
such that $(t,s)$ are a system of global coordinates also for $\Omega_{\ve}$.

\n
Then by proposition \ref{prop1} for every $\ve>0$, the operator $-\Delta^D_{\Omega_{\ve}}$ is unitarily equivalent
to $H_{\ve}$ defined as the closure of the essentially self-adjoint operator $H_{0\ve}$ given by
\begin{equation*}
H_{0\ve}=-
\pd{}{t}\frac{1}{(1+\ve^{\al-1} s\gamma(t/\ve))^2}\pd{}{t}-
 \frac{1}{\ve^{2\al}} \pd{^2}{s^2}+\frac{1}{\ve^2} V_\ve(t,s)\,,
\end{equation*}
\n with
\begin{equation*}
V_\ve(t,s)=-\frac{\gamma(t/\ve)^2}{4(1+\ve^{\al-1}s\gamma(t/\ve))^2}
+\frac{\ve^{\al-1} s\gamma''(t/\ve)}{2(1+\ve^{\al-1}s\gamma(t/\ve))^3}
-\frac{5}{4}\frac{\ve^{2\al -2} s^2\gamma'(t/\ve)^2}{(1+\ve^{\al-1}s\gamma(t/\ve))^4}
\end{equation*}
and
\begin{equation*}
{\mathscr D}(H_{0\ve})=\{
\psi\in L^2(\Omega') \,\, s.t. \,\, \psi\in
C^\infty(\Omega') \,,\,\psi(t,d)=\psi(t,-d)=0 \, , \,
H_{0\ve} \psi \in L^2(\Omega') \} 
\end{equation*}
where we have put $ \Omega'=\RE\times[-d ,d ] $.

\n
The normal modes $\phi_{n}$ satisfy the equation
\begin{equation*}
\left\{\begin{aligned}
&- \frac{1}{\ve^{2\al}} \frac{d^2}{ds^2}\phi_{n}=
\la_{\ve,n} \phi_{n}\\
&\phi_{n}(- d)=\phi_{ n}( d)=0
\end{aligned}\right.\qquad 
n=1,2,3,\dots\,.
\end{equation*}
With
\begin{equation}\label{kingdom}
\la_{\ve,n} = \Big(\frac{n\pi}{2\ve^{\al}d}\Big)^2\, . 
\end{equation}

\n
Let us recall some facts about the low energy behavior of one
dimensional Hamiltonians  (we shall use the results of \cite{BGW}).
We consider the Hamiltonian $\overline{H}$ given by:
\begin{equation}\label{light}
\overline{H} =\overline{H}_0 + \overline{V}\,,\qquad\textrm{with }\;
\oH_0=-\frac{d^2}{dt^2}\,,
\end{equation}
\n where we assume $\int_\erre\oV(t)dt\neq0$ and
$e^{a|\cdot|}\overline V\in L^1(\erre)$ 
for some $a>0$; all the following results hold under these assumptions 
on the potential. Let us denote the free resolvent by 
$G_k=(\overline{H}_0 - k^2 )^{-1}$, its
integral kernel is given by
\begin{equation}\label{Gk}
G_k (t,t') = \frac{i}{2k} e^{ik|t-t'|}\qquad k^2\in\CO\backslash\erre^+,\,\Im k>0\,.
\end{equation}
In order to discuss the low energy behavior of the resolvent $(\overline{H} -k^2)^{-1}$
one reduces the problem to the analysis of the properties of the transition operator $T(k)$
\begin{equation}\label{Tk}
T(k)= (1+ u G_k v)^{-1} \qquad\Im k \geqslant 0, \,k\neq 0, \,k^2\notin \Sigma_p(\overline{H} )
\end{equation}
where we introduced the following two functions
\begin{equation}\label{uv}
v(t)=|\VV(t)|^{1/2}\,,\quad
 u(t)=\sgn[\VV(t)]|\VV(t)|^{1/2}\,.
\end{equation}
and $\Sigma_p(\oH)$ indicates the point spectrum of $\oH$. For
this purpose it is necessary to isolate the singularity of the free
resolvent. In fact we put 
\begin{equation*}
uG_k v = \frac{i}{2k} (v, \cdot ) u + M(k)
\end{equation*}
\n where $(\cdot\,,\cdot)$ indicates the standard scalar product in $L^2(\erre)$. Under our assumptions on $\oV$ the operator $M(k)$ admits the following expansion which converges  in the Hilbert-Schmidt norm
\begin{equation*}
M(k)= \sum_{n=0}^{\infty}(ik)^n  m_n \qquad m_n(t,t') = -1/2 \,u(t) \frac{|t-t'|^{n+1} }{(n+1)!} v(t')\,,\;
n=0,1, \ldots\,.
\end{equation*}
Let us notice that, under our assumptions on $\overline V$,
$(v,u)\neq0$, then we can
define the following two operators
\begin{equation*}
P=\frac{1}{(v,u)}(v,\cdot\,)u\;,\quad Q=1-P
\end{equation*}
such that
\begin{equation*}
T(k)= \lf(1+  \frac{i(v,u)}{2k} P +M(k)\ri)^{-1} \,.
\end{equation*}
We say that $\overline{H}$ has a zero
energy resonance if there exist $\psi_r\in L^{\infty}(\erre)$,  
$\psi_r \notin L^2(\erre)$ such that $\overline{H} \psi_r =0$ in distributional sense; this
is equivalent to the existence of $\varphi_0 \in L^2(\erre)$ such that
\begin{equation}\label{eqphi0}
\varphi_0 + Q m_0 Q \varphi_0=0\,.
\end{equation}
Furthermore, if $\varphi_0$ exists, it is unique, up to a trivial
multiplicative constant, and we can define two constants $c_1$ and $c_2$
by 
\begin{equation}\label{c1c2}
c_1=\frac{(v,m_0\varphi_0)}{(v,u)}\;,\quad 
c_2=\frac{1}{2}((\cdot)v,\varphi_0)\,.
\end{equation}
We can choose $\varphi_0$ such that $c_1$ and $c_2$ are
real. Furthermore under our assumptions on $\oV$ the constants $c_1$
and $c_2$ can not be both zero, in such a case $\psi^r$
would be in $L^2(\erre)$ then zero would be an eigenvalue for $\oH$
(see Lemma 2.2. in \cite{BGW}), but this is impossible under our
assumptions on the potential, see 
Theorem 5.2. in \cite{JN}.

\n
Let $\overline{H}^{r}$ be the following family of self-adjoint operators depending on $c_1$ and $c_2$
\begin{equation*}
{\mathscr D}(\overline{H}^{r}) = \{ f\in H^2(\erre \setminus 0 ) \,\, s.t. \,\,
 (c_1 + c_2 ) f(0^+ ) = (c_1 - c_2 ) f(0^- )
\, , \, (c_1 - c_2 ) f' (0^+ ) = (c_1 + c_2 ) f' (0^- ) \}
\end{equation*}
\begin{equation*}
\overline{H}^{r} f = - \frac{d^2 f}{dt^2} \qquad t\neq 0\,.
\end{equation*}
The Hamiltonian $\overline{H}^{r}$ is a self-adjoint extension of the
symmetric operator $-\Delta$ in dimension one  defined on
$C_0^{\infty}(\erre \setminus \{ 0 \} )$. This kind of point interactions are usually referred to as  ``scale invariant'' (see \cite{HC} and references therein). We refer to \cite{ABD} for
a comprehensive characterization of the point perturbations of the
Laplacian in dimension one (see also, e.g., \cite{AK}).

\n
We denote the one dimensional Laplacian with Dirichlet boundary conditions at the origin by $\overline{H}^{D}$
\begin{equation*}
{\mathscr D}(\overline{H}^{D} ) = \{ f\in H^2(\erre \setminus 0 ) \cap
H^1(\erre  )\,\,s.t.\,\,  f(0)=0 \} 
\end{equation*}
\begin{equation*}
\overline{H}^{D} f = - \frac{d^2 f}{dt^2} \qquad t\neq 0\,.
\end{equation*}

\n
We  want to discuss the convergence of the resolvent of
$-\Delta^D_{\Omega_{\ve}}$, as $\ve \to 0$, to a one dimensional
operator on the broken line. Since the proposition
\ref{prop1} holds  we can reduce the problem to the analysis of the
convergence of $H_\ve$. 

\n The normal
modes $\phi_{n}$ diagonalize the transversal part of the kinetic
term in $H_\ve$, then they provide a useful reference frame for
discussing the limit $\ve \to 0$. For these reason we shall consider
the matrix elements of the resolvent of $H_\ve$ with respect to $\phi_{n}$ 
and $\phi_{m}$ and we shall discuss the limit of these
operators. 

\n The term $\ve^{-\alpha}$ in the definition of $\lambda_{\ve,n}$, see
formula \eqref{kingdom}, indicates that the transversal part of the
kinetic term of $H_{\ve}$ is divergent. In 
order 
to obtain a non trivial limit, following a standard procedure
(see, e.g., \cite{DT} and \cite{Po}), we regularize the resolvent of
$H_{\ve}$ by subtracting the divergent
eigenvalue $\lambda_{\ve,n}$ from the spectral parameter. We shall prove that only the diagonal
elements survive as $\ve
\to 0$.

\n
Under our hypothesis the resolvent  of $H_{\ve}$ admits the integral kernel 
$( H_{\ve} - k^2 - \la_{\ve,m} )^{-1} (t,s,t',s')$, see Theorem II.37 in \cite{Si}, and let us 
introduce the operator $R_{n,m}^{\ve}(k^2) : L^2(\erre) \rightarrow L^2(\erre) $ whose
integral kernel is given by
\begin{equation*}
R_{n,m}^{\ve}(k^2, t,t') =  \int_{-d}^{d}ds\,ds' \,
\phi_{n}(s)  ( H_{\ve} - k^2 - \la_{\ve,m} )^{-1}(t,s,t',s')
\phi_{m} (s')\,.
\end{equation*}
It is straightforward to notice that $R_{n,m}^{\ve}(k^2)$ are bounded operator valued
analytic functions of $k^2$, for $k^2\in\CO\backslash\RE$ and $\Im k>0$.

\n
Let us fix some notation: for a given open set $E \subset \erre^n$ and $p\geq1$
we denote the norm of $L^p( E)$ by 
$\| \cdot \|_{L^p(E)}$, when $E$ is omitted it is understood $E=\erre$, furthermore we denote the
Banach space of bounded operators from $L^p(E)$ to $L^q(E)$, $p,q\geq1$,
equipped with its natural norm by
${\mathscr B}(L^p(E),L^q(E) )$; we also denote the Hilbert-Schmidt norm for operators in 
${\mathscr B}(L^2,L^2 )$ by $\|\cdot \|_{HS}$. We indicate with $c$ a numerical constant whose value may
change from line to line. Moreover, we denote by $\ulim_{\ve\to0} $ the limit in the uniform topology
of $\mathscr B(L^2,L^2)$. 

\n
Now we can state our main theorem.
\begin{theorem}\label{mainth}
Assume that $\Ga$ has no self-intersections and that
$\ga $ is piecewise $C^2$, has compact support and $\ga',\ga''$
are bounded. Moreover
take $\al > 5/2$ and put $\overline{V} = - \ga^2/4$. Then two cases
can occur:
 
\n 1) There does not exist a zero energy resonance for the
Hamiltonian $\overline{H} $. In such a case
\begin{equation*}
\ulim_{\ve \rightarrow 0} R_{n,m}^{\ve}(k^2) = \delta_{n,m}( \overline{H}^{D}-k^2
)^{-1} \qquad k^2\in\CO\backslash\erre,\;\Im k>0\,.
\end{equation*}

\n 2) There exists a zero energy resonance for the Hamiltonian
$\overline{H} $. In such a case
\begin{equation*}
\ulim_{\ve \rightarrow 0} R_{n,m}^{\ve}(k^2) = \delta_{n,m}(
\overline{H}^{r}-k^2 )^{-1} 
 \qquad k^2\in\CO\backslash\erre,\;\Im k>0\,.
\end{equation*}
Here $\delta_{n,m}$ indicates the Kronecker symbol, i.e. $\delta_{n,m}=0$ if $n\neq m$ and $\delta_{n,n}=1$.
\end{theorem}

\n We shall prove theorem \ref{mainth} in the next section.

\n
The assumptions of theorem \ref{mainth} are not optimal: one could require
that $\ga$ has a suitable decay at infinity, as for instance in \cite{DE} and
\cite{ESjmp}, where it is assumed that $\ga$ belongs to some weighted $L^p$,
instead of compact support, but we are not interested in the maximal
generality.

\section{\label{sec2}Proof of Theorem \ref{mainth}}

\n
In this section we shall prove theorem \ref{mainth}; 
first we shall prove three lemmas and then the proof of theorem \ref{mainth}
will immediately follow.

\n 
We are interested in the limit of the following operator for $\ve\to0$ 
\begin{equation*}
\overline H_\ve=\overline H_0+\frac{1}{\ve^2}\overline V(\cdot/\ve)
=-\frac{d^2}{dt^2}+\frac{1}{\ve^2}\overline V(t/\ve)\,.
\end{equation*}
\n Before stating our result on the convergence of $\overline H_\ve$,
let us introduce some notation and spend  few words on the
correspondence between our problem and the low energy expansion of the
resolvent $(\overline H-k^2)^{-1}$.

\n Let us assume $k^2\in\CO\backslash\erre$ and $\Im k>0$. Define the dilation operator $U_\ve$
\begin{equation*}
(U_\ve f)(t)=\frac{1}{\ve^{1/2}}f(t/\ve)\,,
\end{equation*}
\n the operator $U_\ve$ is unitary on $L^2(\erre)$ and, by using
the identity $\oH_\ve=\ve^{-2}U_\ve\oH U_\ve^*$, one obtains
\begin{equation}\label{quo}
(\oH_\ve-k^2)^{-1}=\ve^{2}U_\ve(\oH-\ve^2k^2)^{-1}U_\ve^*\,.
\end{equation} 
\n Here $^*$ indicates the adjoint. The resolvent of $\oH$ can be written as 
\begin{equation}\label{qui}
(\oH-k^2)^{-1}=G_k-G_kvT(k)uG_k
\end{equation}
\n where $G_k$, $T(k)$, $u$ and $v$ were defined in (\ref{Gk}), (\ref{Tk})
and (\ref{uv}) respectively. By using equation $(\ref{qui})$ in
$(\ref{quo})$ one obtains the 
following formula for $(\oH_\ve-k^2)^{-1}$ 
\begin{equation}\label{resolvent}
(\overline H_\ve-k^2)^{-1}=G_k-\frac{1}{\ve}A_\ve(k)T(\ve
k)C_\ve(k)
\end{equation}
\n where $A_\ve(k)$ and $C_\ve(k)$ are
defined via their integral kernels
\begin{align*}
&A_\ve(k;t,t')=G_k(t-\ve t')v(t')\\
&C_\ve(k;t,t')=u(t)G_k(\ve t-t')\,.
\end{align*}  
\n To obtain the limit of the resolvent (\ref{resolvent}) we shall use
the results of \cite{BGW} about the low energy behavior of
$T(k)$, we recall such results in the following
\begin{proposition}\label{prop2}Let $\int_{\erre}
\overline{V}(t) dt \neq 0$ and $e^{a|\cdot|}\overline V\in L^1(\erre)$
for some $a>0$. Then two cases can occur:\\

\n 1) There does not exist a zero energy resonance for the Hamiltonian
$\overline H$. In such a case no solution, $\varphi_0\in L^2(\erre)$, of equation
(\ref{eqphi0}) exists.\\

\n 2) There exists a zero energy resonance for the Hamiltonian $\overline
H$. In such a case a solution, $\varphi_0\in L^2(\erre)$, of equation (\ref{eqphi0})
exists and the constants $c_1$ and $c_2$ defined in (\ref{c1c2}) do
not vanish simultaneously.\\

\n The operator $T(k)$ has the 
following norm convergent series  expansion around $k=0$
\begin{equation*}
T(k)=\sum_{n=p}^\infty (i k)^nt_n
\end{equation*}
\n with $p=0$ in case 1 and $p=-1$ in case 2.
\end{proposition}

\n For the proof of proposition \ref{prop2} we refer to
\cite{BGW}. There the authors give also some recursive formulas to get all the 
terms $t_j$ of the
series expansion.\begin{flushright} $\square$\end{flushright}

\n Now we can state and prove the following lemma on the convergence
of $(\oH_\ve-k^2)^{-1}$:
\begin{lemma}\label{lemma1}Let $\int_{\erre}
\overline{V}(t) dt \neq 0$ and $e^{a|\cdot|}\overline V\in L^1(\erre)$
for some $a>0$. Then two cases can occur:\\

\n 1) There does not exist a zero energy resonance for the Hamiltonian
$\overline H$. In such a case 
\begin{equation*}
\ulim_{\ve\to0}
(\overline H_\ve-k^2)^{-1}=(\overline H^D-k^2)^{-1}
\qquad k^2\in\CO\backslash\erre,\,\Im k>0\,.
\end{equation*}

\n 2) There exists a zero energy resonance for the Hamiltonian $\overline
H$. In such a case 
\begin{equation*}
\ulim_{\ve\to0}
(\overline H_\ve-k^2)^{-1}=(\overline H^r-k^2)^{-1}
\qquad k^2\in\CO\backslash\erre,\,\Im k>0
\end{equation*}
\end{lemma}
\begin{proof}
\n Let us consider first the case 2. Under the assumptions on $\oV(t)$ and for $\Im k>0$ the operators
$A_\ve(k)$ and $C_\ve(k)$ are Hilbert-Schmidt and
\begin{align}
&A_\ve(k;t,t')= \Big(\frac{i}{2k}e^{ik|t|}-\frac{1}{2}e^{ik|t|}(|t-\ve
t'|-|t|)+a_\ve(k;t,t')\Big)v(t')
\label{Aveseries}
\\
&C_\ve(k;t,t')=
u(t)\Big(\frac{i}{2k}e^{ik|t'|}-\frac{1}{2}e^{ik|t'|}(|\ve t-
t'|-|t'|)+c_\ve(k;t,t')\Big)
\label{Cveseries}
\end{align}
\n where 
\begin{align}
&a_\ve(k;t,t')=
-\frac{ik}{2}e^{ik|t|}\int_0^{|t-\ve t'|-|t|}
e^{ik\tau}(|t-\ve t'|-|t|-\tau)d\tau 
\label{hope}
\\ 
&c_\ve(k;t,t')=
-\frac{ik}{2}e^{ik|t'|}\int_0^{|\ve t-t'|-|t'|}
e^{ik\tau}(|\ve t-t'|-|t'|-\tau)d\tau\,.
\end{align}
\n The following estimates for $a_\ve(k)$ and $c_\ve(k)$ hold
\begin{align*}
&\|a_\ve(k)v\|_{HS}\leqslant
\frac{|k|}{4}\frac{1}{\sqrt{\Im k}}\|(\cdot)^2v\|_{L^2}\,\ve^2\\
&\|uc_\ve(k)\|_{HS}\leqslant
\frac{|k|}{4}\frac{1}{\sqrt{\Im k}}\|(\cdot)^2u\|_{L^2}\,\ve^2\,.
\end{align*}
\n From proposition \ref{prop2} it follows that
\begin{equation}
T(\ve k)=\frac{1}{ik\ve}t_{-1}+t_0+ik\ve t_1+b_\ve(k)\label{Tveseries}
\end{equation}
\n with $\|b_\ve(k)\|_{\mathscr B(L^2,L^2)}\leqslant c\ve^2$. From (\ref{Aveseries}), (\ref{Cveseries})
and (\ref{Tveseries}) we obtain the following formula for  the
 integral kernel of $A_\ve(k)T(\ve
k)C_\ve(k)$
\begin{equation}\label{stone}
\begin{aligned}
\big(A_\ve(k)T(\ve
k)C_\ve(k)\big)(t,t')=
\int d\tau d\tau'\Bigg[&\Big(\frac{i}{2k}e^{ik|t|}-\frac{1}{2}e^{ik|t|}(|t-\ve
\tau|-|t|)\Big)\times\\
&\times v(\tau)
\Big(\frac{1}{ik\ve}t_{-1}(\tau,\tau')+t_0(\tau,\tau')+ik\ve t_1(\tau,\tau')\Big)
u(\tau')\times\\
&\times\Big(\frac{i}{2k}e^{ik|t'|}-\frac{1}{2}e^{ik|t'|}(|\ve \tau'-
t'|-|t'|)\Big)\Bigg]+r_{\ve1}(k;t,t')
\end{aligned}
\end{equation}
\n with $\|r_{\ve1}(k)\|_{\mathscr B(L^2,L^2)}\leqslant c\ve^2$. We shall use the following properties of $t_{-1}$, $t_0$ and 
$t_1$
\begin{gather}
t_{-1}u=0\,;\quad t_{-1}^*v=0\,;\quad(v,t_0u)=0\,;\label{eqBGW1}
\\
((\cdot)v,t_{-1}u(\cdot))=\frac{2c_2^2}{c_1^2+c_2^2}\,;\quad
((\cdot)v,t_0u)=(v,t_0u(\cdot))=\frac{2c_1c_2}{c_1^2+c_2^2}\,;\quad
(v,t_1u)=-\frac{2c_2^2}{c_1^2+c_2^2}\,.\label{eqBGW2}
\end{gather}
\n For a detailed derivation of $(\ref{eqBGW1})$ and
$(\ref{eqBGW2})$ we refer to \cite{BGW}. Let us state the following equality
\begin{equation}\label{modulo1}
|t-\ve \tau|-|t|=-\ve \tau\sgn(t)+2(\ve \tau-t)\XX_{[0,\ve
  \tau]}(t)\Theta(\tau)
+2(t-\ve \tau)\XX_{[\ve \tau,0]}(t)\Theta(-\tau)
\end{equation}
\n where $\XX_{[a,b]}(t)$ is the characteristic function of the
interval $[a,b]$ and $\Theta(\tau)$ is the Heaviside function. The estimates
\begin{align}
&\Bigg( \int dtd\tau
\Big|e^{ik|t|}(\ve\tau-t)\XX_{[0,\ve\tau]}(t)\Theta(\tau)v(\tau)\Big|^2
\Bigg)^{1/2}\leqslant
\ve^{3/2}\|(\cdot)^{3/2}v\|_{L^2}
\label{estimate1}
\\
&\Bigg( \int dtd\tau
\Big|e^{ik|t|}(t-\ve\tau)\XX_{[\ve\tau,0]}(t)\Theta(-\tau)v(\tau)\Big|^2
\Bigg)^{1/2}\leqslant
\ve^{3/2}\|(\cdot)^{3/2}v\|_{L^2}\,.
\label{estimate2}
\end{align}
\n  hold. By using
the equality (\ref{modulo1}) and the estimates (\ref{estimate1}) and
(\ref{estimate2}), and the corresponding ones for the term
$u(\tau')e^{ik|t'|}(|\ve \tau'-t'|-|t'|)$, in 
(\ref{stone}) and taking into account  equations (\ref{eqBGW1}) and
(\ref{eqBGW2}) we obtain 
\begin{equation}
\begin{aligned}
\label{floor}
\big(A_\ve(k)T(\ve
k)C_\ve(k)\big)(t,t')=&\ve\Bigg(-
2ik\frac{c_2^2}{c_1^2+c_2^2}G_k(t)G_k(t')
+\frac{2}{ik}\frac{c_2^2}{c_1^2+c_2^2}G_k'(t)G_k'(t')+\\
&-2\frac{c_1c_2}{c_1^2+c_2^2}G_k(t)G_k'(t')
-2\frac{c_1c_2}{c_1^2+c_2^2}G_k'(t)G_k(t')\Bigg)+r_{\ve2}(k;t,t')
\end{aligned}
\end{equation}
with $\|r_{\ve2}\|_{\mathscr B(L^2,L^2)}\leqslant c\ve^{3/2}$. Here $G_k'(t)$ is the derivative of $G_k(t)$
\begin{equation*}
G_k'(t)=
-\frac{\sgn(t)}{2}e^{ik|t|}\qquad k^2\in\CO\backslash\erre^+,\,\Im
k>0\,. 
\end{equation*} 
Then from (\ref{resolvent}) and (\ref{floor}) it follows that
\begin{equation*}
\ulim_{\ve\to0}(\overline H_\ve-k^2)^{-1}=\overline
 R^r(k^2)\qquad k^2\in\CO\backslash\erre,\,\Im k>0\,,
\end{equation*}
with
\begin{equation}
\begin{aligned}
\label{resHc1c2}
\overline R^r(k^2;t,t')=&G_k(t-t')+
2ik\frac{c_2^2}{c_1^2+c_2^2}G_k(t)G_k(t')
-\frac{2}{ik}\frac{c_2^2}{c_1^2+c_2^2}G_k'(t)G_k'(t')+\\
&+2\frac{c_1c_2}{c_1^2+c_2^2}G_k(t)G_k'(t')
+2\frac{c_1c_2}{c_1^2+c_2^2}G_k'(t)G_k(t')\,.
\end{aligned}
\end{equation} 

\n We need to prove that the operator $\overline{R}^r(k^2)$ is the resolvent of
the Hamiltonian $\oH^r$; in facts we shall prove that $\overline{R}^r(k^2)$ is the resolvent of
an operator which is a self-adjoint extension of the Laplacian defined on $C^{\infty}_0 (\erre \setminus \{ 0\} )$
and which satisfies the same boundary conditions as $\overline H^r$ at
the origin. 

\n
A tedious but straightforward calculation, based on the fact
that $G_k(t-t')$ satisfies the resolvent identity, shows that 
\begin{equation*}
\overline R^r(k^2)-
\overline R^r(p^2)=(k^2-p^2)
\overline R^r(p^2)\overline R^r(k^2)
\qquad
k^2,p^2\in\CO\backslash\erre,\,\Im k>0,\,\Im p>0\,.
\end{equation*}
\n Moreover
\begin{equation*}
\big(\overline R^r(z)\big)^*=\overline
R^r(\bar z)\qquad z\in\CO\backslash\erre,\,\Im\sqrt{z}>0
\end{equation*}
\n where $^-$ indicates the complex
conjugation. The operator $\overline R^r(k^2)$ is injective
because $G_k\notin H^2(\erre)$ and $G_k'\notin H^2(\erre)$, then it
is invertible on $\textrm{Ran}[\overline{R}^{r}(k^2)]$ and defines a symmetric operator with domain
$\textrm{Ran}[\overline 
R^r(k^2)]$.  Let us define the function $g_f(t)=\big(\overline
R^r(k^2)f\big)(t)$, with $f\in L^2(\erre)$,
$k^2\in\CO\backslash\erre$, $\Im k>0$:
\begin{equation*}
\begin{aligned}
g_f(t)=&
\big(G_kf\big)(t)+
2ik\frac{c_2^2}{c_1^2+c_2^2}G_k(t)\big(G_kf\big)(0)
+\frac{2}{ik}\frac{c_2^2}{c_1^2+c_2^2}G_k'(t)\big(G_k'f\big)(0)+\\
&-2\frac{c_1c_2}{c_1^2+c_2^2}G_k(t)\big(G_k'f\big)(0)
+2\frac{c_1c_2}{c_1^2+c_2^2}G_k'(t)\big(G_kf\big)(0)
\end{aligned}
\end{equation*}
\n  where
we used $\int_\erre G_k'(\tau)f(\tau)d\tau=-\big(G_k'f\big)(0)$. A direct
computation gives
\begin{equation}
\label{machine}
g_f(0^+)=(c_1-c_2)K_f\,;\quad g_f(0^-)=(c_1+c_2)K_f\;;\qquad
g_f'(0^+)=(c_1+c_2)K_f'\,;\quad g_f'(0^-)=(c_1-c_2)K_f'
\end{equation}
\n with 
\begin{equation*}
K_f=\Big(\frac{c_1}{c_1^2+c_2^2}\big(G_kf\big)(0)-
\frac{i}{k}\frac{c_2}{c_1^2+c_2^2}\big(G_k'f\big)(0)\Big)\,;\qquad
K_f'=\Big(-ik\frac{c_2}{c_1^2+c_2^2}\big(G_kf\big)(0)+
\frac{c_1}{c_1^2+c_2^2}\big(G_k'f\big)(0)\Big)\,.
\end{equation*}
\n Conditions (\ref{machine}) are equivalent to 
\begin{equation*}
(c_1+c_2)g_f(0^+)=(c_1-c_2)g_f(0^-)\;;\qquad
(c_1-c_2)g_f'(0^+)=(c_1+c_2)g_f'(0^-)\,,
\end{equation*}
\n then $\textrm{Ran}[\overline
R^r(k^2)]=\mathscr D(\overline H^r)$. Moreover if
$g_f(t)=\big(\overline R^r(k^2)f\big)(t)$ is such that
$g_f(0^+)=g_f(0^-)=g_f'(0^+)=g_f'(0^-)=0$, then equations
\eqref{machine} together with the definitions of $K_f$ and $K_f'$ give
$(G_kf)(0)=(G_k'f)(0)=0$,  from which 
 $g_f(t)=(G_kf)(t)$. This means that the operator with resolvent
$\overline R^r(k^2)$ acts as the Laplacian on functions with support that does
not contain the origin, since its domain coincides with $\mathscr
D(\oH^r)$ it coincides with $\oH^r$.

\n Alternatively, since $\overline
H^r$ is a 
self-adjoint extension of the operator $-\Delta$ with
$\mathscr D(-\Delta)=C_0^\infty(\erre\backslash0)$, one can prove that
$\overline R^r(k^2)$ is  the integral
kernel of the  resolvent of $H^r$  by using the results of \cite{ABD}
or the
``modified Krein's resolvent formula'' derived in
\cite{AP} (see also \cite{P}).\\

\n The proof of the case 1 is analogous. If there is not a zero energy
resonance the series expansion of $T(\ve k)$ starts from the order
zero in $\ve$, and  the following equations replace the ones in
$(\ref{eqBGW1})$ and $(\ref{eqBGW2})$
\begin{equation}\label{clocks}
(v,t_0u)=0\,;\quad((\cdot)v,t_0u)=(v,t_0u(\cdot))=0\,;\quad
(v,t_1u)=-2\,.
\end{equation}
\n Then in such a case the following expansion holds
\begin{equation*}
\begin{aligned}
\big(A_\ve(k)T(\ve
k)C_\ve(k)\big)(t,t')=-2ik\ve
\frac{i}{2k}e^{ik|t|}\frac{i}{2k}e^{ik|t'|}+r_{\ve 3}(k;t,t')\,,\quad
\textrm{with }
\|r_{\ve3}(k)\|_{\mathscr B(L^2,L^2)}\leqslant c\ve^{3/2}
\end{aligned}
\end{equation*}
and 
\begin{equation*}
\ulim_{\ve\to0}(\overline H_\ve-k^2)^{-1}=\overline
 R^{D}(k^2)\qquad k^2\in\CO\backslash\erre,\,\Im k>0
\end{equation*}
where
\begin{equation*}
\overline R^{D}(k^2)=G_k(t-t')+2ikG_k(t)G_k(t')\
\end{equation*}
\n $\overline R^D(k^2)$ is the resolvent of $\oH^D$, and the proof of
the lemma is complete.
\end{proof}

\n
Let us prove two technical estimates that will be used in lemma \ref{lemma3}:
\begin{lemma}Let $\int_{\erre}
\overline{V}(t) dt \neq 0$ and $e^{a|\cdot|}\overline V\in L^1(\erre)$
for some $a>0$ then
\begin{equation}
\limsup_{\ve \to 0} \ve^{1/2} \lf\| \frac{d}{dt}( \overline{H}_\ve - k^2 )^{-1} \ri\|_{{\mathscr B}(L^2,L^2)} 
\leqslant c 
\label{tacchino1}
\end{equation}
\begin{equation}
\limsup_{\ve \to 0}\|(\overline{H}_\ve-k^2)^{-1}\|_{{\mathscr B}(L^2,L^{\infty})} 
\leqslant c 
\label{tacchino2}
\end{equation}
\end{lemma}
\begin{proof}

\n We shall first prove the estimate (\ref{tacchino1}). We use formula
(\ref{resolvent}) and remark that   the
derivative of the
resolvent of the free Laplacian, $G_k'$, is bounded in
$\mathscr B(L^2,L^2)$. Let us consider the
derivative of $A_\ve(k)T(\ve k)C_\ve(k)$. The case with resonance and
the case without resonance must be discussed separately. Let us
assume that $\oH$ has a zero energy resonance, by using equations
(\ref{Aveseries}), (\ref{Cveseries}) and (\ref{Tveseries}) we obtain
\begin{equation*}
\begin{aligned}
\pd{}{t}\big(A_\ve(k)T(\ve
k)C_\ve(k)\big)(t,t')=
\int d\tau d\tau'\Bigg[&
ik\sgn(t-\ve\tau)\Big(\frac{i}{2k}e^{ik|t|}-\frac{1}{2}e^{ik|t|}(|t-\ve
\tau|-|t|)+a_\ve(k;t,\tau)\Big)\times\\
&\times v(\tau)
\Big(\frac{1}{ik\ve}t_{-1}(\tau,\tau')+t_0(\tau,\tau')+ik\ve
t_1(\tau,\tau')
+b_\ve(k;t,\tau)\Big)
u(\tau')\times\\
&\times\Big(\frac{i}{2k}e^{ik|t'|}-\frac{1}{2}e^{ik|t'|}(|\ve \tau'-
t'|-|t'|)+c_\ve(k;t,\tau)\Big)\Bigg]\,.
\end{aligned}
\end{equation*}
\n Following what was done in the lemma \ref{lemma1} we use the
identity (\ref{modulo1}) and the properties of operators $t_{-1}$,
$t_0$ and $t_1$, see (\ref{eqBGW1}) and (\ref{eqBGW2}), to obtain
\begin{equation*}
\begin{aligned}
\pd{}{t}\big(A_\ve(k)T(\ve
k)C_\ve(k)\big)(t,t')=&
\frac{i}{4k}e^{ik|t|}\int\sgn(t-\ve\tau)v(\tau)
t_{-1}(\tau,\tau')u(\tau')\tau'
d\tau d\tau'\,\sgn(t')e^{ik|t'|}+\\
&-\frac{i}{4k}e^{ik|t|}\int\sgn(t-\ve\tau)v(\tau)
t_0(\tau,\tau')u(\tau')d\tau d\tau'\,e^{ik|t'|}
+r_{\ve 4}(k;t,t')
\end{aligned}
\end{equation*}
\n where, for $\ve$ small enough, $\|r_{\ve 4}(k)\|_{\mathscr
  B(L^2,L^2)}\leqslant c\ve$. Now we use the following expression for
the function  $\sgn(t-\ve\tau)$
\begin{equation*}
\sgn(t-\ve\tau)=
\sgn(t)-2\XX_{[0,\ve\tau]}(t)\Theta(\tau)+2\XX_{[\ve\tau,0]}(-\tau)\,.
\end{equation*}
\n Since
$t_{-1}^*v=0$, $(v,t_0u)=0$ and  
\begin{align*}
&\Bigg( \int dtd\tau
\Big|e^{ik|t|}\XX_{[0,\ve\tau]}(t)\Theta(\tau)v(\tau)\Big|^2
\Bigg)^{1/2}\leqslant
\ve^{1/2}\|(\cdot)^{1/2}v\|_{L^2}\\
&\Bigg( \int dtd\tau
\Big|e^{ik|t|}\XX_{[\ve\tau,0]}(t)\Theta(-\tau)v(\tau)\Big|^2
\Bigg)^{1/2}\leqslant
\ve^{1/2}\|(\cdot)^{1/2}v\|_{L^2}\,,
\end{align*}
\n for $\ve$ small enough the  estimate
\begin{equation*}
\Big\|\frac{d}{dt}A_\ve(k)T(\ve
k)C_\ve(k)\Big\|_{\mathscr B(L^2,L^2)}\leqslant
c\ve^{1/2}
\end{equation*}
\n holds,  from which the limit estimate  (\ref{tacchino1}) immediately
follows. The case without resonance is analogous. The only difference
is in the series expansion of $T(\ve k)$, in fact the series starts
with the term of order zero in $\ve$ and equations \eqref{eqBGW1} and
\eqref{eqBGW2} are replaced by \eqref{clocks}.\\
 
\n Let us prove the estimate (\ref{tacchino2}). We use again the resolvent formula
(\ref{resolvent}). The  resolvent of the free Laplacian, $G_k$ satisfies
\begin{equation*}
\|G_k\|_{\mathscr B(L^2,L^\infty)}\leqslant \frac{1}{2|k|\sqrt{\Im k}}\,.
\end{equation*}

\n Let us set $A_\ve(k)=A_{1,\ve}(k)+A_{2,\ve}(k)$, with 
\begin{align*}
&A_{1,\ve}(k;t,t')=\frac{i}{2k}e^{ik|t|}v(t')\\
&A_{2,\ve}(k;t,t')=-\frac{1}{2}e^{ik|t|}(|t-\ve
t'|-|t|)v(t')+a_\ve(k;t,t')v(t')\,,
\end{align*}
\n where $a_\ve(k;t,t')$ was given in (\ref{hope}). A direct computation yields 
\begin{equation}\label{hail}
\|A_{2,\ve}(k)\|_{\mathscr B(L^2,L^\infty)}\leqslant 
\frac{1}{2}\|(\cdot)v\|_{L^2}\,\ve+
\frac{|k|}{4}
\|(\cdot)^2v\|_{L^2}\,\ve^2\,.
\end{equation}

\n Moreover the following limit for the $\mathscr B(L^2,L^2)$-norm of
the operator $T(\ve k)C_\ve(k)$ holds
\begin{equation}\label{giornata}
\lim_{\ve\to0}\|T(\ve k)C_\ve(k)\|_{\mathscr B(L^2,L^2)}\leqslant c\,.
\end{equation}
\n In fact, if there does not exist  a zero energy resonance the limit (\ref{giornata})
is a consequence of the fact that $\lim_{\ve\to0}\|T(\ve
k)\|_{\mathscr B(L^2,L^2)}\leqslant c$ and
$\lim_{\ve\to0}\|C_\ve(k)\|_{\mathscr B(L^2,L^2)}\leqslant c$. If there is
a zero energy resonance  we
can use (\ref{Cveseries}) and (\ref{Tveseries}). In such a
case equation (\ref{giornata}) is a consequence of the fact that 
$t_{-1}u=0$ and $\big||\ve t-t'|-|t'|\big|\leqslant\ve|t|$. From
(\ref{hail}) and (\ref{giornata}) we obtain
\begin{equation*}
\limsup_{\ve\to0}\frac{1}{\ve}
\|A_{2,\ve}(k)T(\ve k)C_\ve(k)\|_{\mathscr B(L^2,L^\infty)}
\leqslant c\,.
\end{equation*}
\n The limit of $\ve^{-1}\|A_{1,\ve}(k)T(\ve k)C_\ve(k)\|_{\mathscr
  B(L^2,L^\infty)}$, as $\ve\to0$,  can be studied as follows. In the presence of a zero energy
resonance the integral kernel of
$A_{1,\ve}(k)T(\ve k)C_\ve(k)$ is 
\begin{equation*}
\begin{aligned}
\big(A_{1,\ve}(k)T(\ve
k)C_\ve(k)\big)(t,t')=
\frac{i}{2k}e^{ik|t|}\int d\tau d\tau'\Bigg[&v(\tau)
\Big(\frac{1}{ik\ve}t_{-1}(\tau,\tau')+t_0(\tau,\tau')+
ik\ve t_1(\tau,\tau')+b_\ve(k;\tau,\tau')\Big)
u(\tau')\times\\
&\times\Big(\frac{i}{2k}e^{ik|t'|}-\frac{1}{2}e^{ik|t'|}(|\ve \tau'-
t'|-|t'|)+c_\ve(k;\tau',t')\Big)\Bigg]
\end{aligned}
\end{equation*}
\n The modulus of the integral is of order $\ve$. This statement can
be proved by reiterating what was done in lemma \ref{lemma1}, for this
reason we  do not give the details of the proof. The term
with $|\ve \tau'-t'|-|t'|$ can be rewritten by using formula
(\ref{modulo1}). Then the properties (\ref{eqBGW1}) and (\ref{eqBGW2}) of the 
operators $t_{-1}$, $t_0$ and $t_1$ can be used to evaluate the term
of order $\ve$. The modulus of the remainder is of order $\ve^{3/2}$
because $\|c_\ve(k)\|_{\mathscr B(L^2,L^2)}\leqslant c\ve^2$,
$\|b_\ve(k)\|_{\mathscr B(L^2,L^2)}\leqslant c\ve^2$ and because the estimates
(\ref{estimate1}) and (\ref{estimate2}) hold. Then
\begin{equation}\label{scientist}
\limsup_{\ve\to0}\frac{1}{\ve}
\|A_{1,\ve}(k)T(\ve k)C_\ve(k)\|_{\mathscr B(L^2,L^\infty)}
\leqslant c\,,
\end{equation}
\n and the estimate (\ref{tacchino2}) immediately follows. If there
does not exist a zero energy resonance, the proof of (\ref{scientist}) is
analogous, but the series expansion of $T(\ve k)$ starts with the term
of order zero in $\ve$ and instead of (\ref{eqBGW1}) and
(\ref{eqBGW2}), equations (\ref{clocks}) hold.
\end{proof}

\n
Let us introduce the operator $\Hboh_{\ve}$ defined as the closure of the
essentially self-adjoint
operator $\Hboh_{0\ve}$:
\begin{equation*}
\Hboh_{0\ve}= -
\pd{^2}{t^2}-
\frac{1}{\ve^{2\al}}  \pd{^2}{s^2}+\frac{1}{\ve^2} \overline{V}(t/\ve)\,,
\end{equation*}
\n 
and
\begin{equation*}
{\mathscr D}(\Hboh_{0\ve})=\{\psi\in L^2(\Omega' )
 \,\, s.t. \,\,
\psi\in C^\infty(\Omega' )\,,\,
\psi(t,d)=\psi(t,-d)=0 \, , \,
\Hboh_{0\ve} \psi \in L^2(\Omega') \}\,.
\end{equation*}
\n Consider the matrix elements $\Rboh_{n,m} $ with respect to  the
normal modes $\phi_{n}$ and  $\phi_{m}$: 
\begin{equation*}
\Rboh_{n,m} (k^2;t,t') =  \int_{-d}^{d}ds\,ds' 
\phi_{n}(s)  ( \Hboh_{\ve} - k^2 - \la_{\ve,m} )^{-1}(t,s,t',s')
\phi_{m} (s')
\end{equation*}

\n
Notice that $\Rboh_{n,m} (k^2) = \de_{n,m} ( \overline{H}_{\ve} - k^2 )^{-1}$ since 
$ \Hboh_{\ve}$ is a separable Hamiltonian.
\begin{lemma}
Assume that $\Ga$ has no self-intersections,
$\ga $ is piecewise $C^2$, has compact support and that $\ga',\ga''$
are bounded. Moreover  take $\al > 5/2$ and
put $\overline{V}= -\ga^2/4$. Then
\begin{equation}
\ulim_{\ve \rightarrow 0}  \big(R_{n,m}^{\ve}(k^2) -\Rboh_{n,m}
(k^2)\big)  = 0
\qquad k^2\in\CO\backslash\RE,\;\Im k>0\,.
\label{propano}
\end{equation} 
\label{lemma3}
\end{lemma}
\begin{proof}
In order to prove \eqref{propano} it is sufficient to prove 
\begin{equation}
 \lf| \lf( g, 
\lf(
R_{n,m}^{\ve}(k^2) - \Rboh_{n,m} (k^2) 
\ri) f
\ri) \ri| \leqslant c \ve^{\al - 5/2}\|  g\|_{L^2} \|  f\|_{L^2}
\label{mulo}
\end{equation}
for any $f,g \in C_0^{\infty}$. Using the resolvent identity we have
\begin{equation*}
\begin{aligned}
&( H_{\ve} - k^2 - \la_{\ve,m} )^{-1} - (\Hboh_{\ve} - k^2 - \la_{\ve,m} )^{-1} =\\
&=( H_{\ve} - k^2 - \la_{\ve,m} )^{-1} 
\lf[ 
\ve^{\al -2} b\lf( \frac{\cdot}{\ve}, \cdot\ri) \pd{}{t}+ \frac{1}{\ve^2} \lf( V\lf( \frac{\cdot}{\ve}, \cdot\ri) - 
\overline{V}\lf(\frac{\cdot}{\ve}\ri) \ri) 
\ri]
(\Hboh_{\ve}- k^2 - \la_{\ve,m} )^{-1}
\end{aligned}
\end{equation*}
with $b(t,s) = -2 s \ga'(t) / (1 + \ve^{\al -1} s \ga(t) )^3 $. 
Therefore it is sufficient to estimate $I_1$ and $I_2$ given by 
\begin{equation*}
I_1 =
\lf( g \otimes \phi_{n},
( H_{\ve} - k^2 - \la_{\ve,m} )^{-1}
\ve^{\al -2} b\lf( \frac{\cdot}{\ve}, \cdot\ri) \pd{}{t}
(\Hboh_{\ve} - k^2 - \la_{\ve,m} )^{-1}f \otimes \phi_{m}\ri)
\end{equation*}
\begin{equation*}
I_2 =
\lf( g \otimes \phi_{n},
( H_{\ve} - k^2 - \la_{\ve,m} )^{-1}
\frac{1}{\ve^2} \lf( V\lf( \frac{\cdot}{\ve}, \cdot\ri) -
\overline{V}\lf(\frac{\cdot}{\ve}\ri) \ri) 
(\Hboh_{\ve} - k^2 - \la_{\ve,m} )^{-1}f \otimes \phi_{m}\ri)
\end{equation*}
Since $H_{\ve}$ is a separable Hamiltonian we have:
\begin{equation*}
(\Hboh_{\ve} - k^2 - \la_{\ve,m} )^{-1}f \otimes \phi_{m} =
\big(( \overline{H}_\ve - k^2 )^{-1} f\big) \otimes \phi_{m}
\end{equation*}

\n
Let us discuss $I_1$: using \eqref{tacchino1}, Cauchy-Schwarz inequality and the estimate 
\begin{equation}
 \|( H_{\ve} - k^2 - \la_{\ve,m} )^{-1}
 \|_{{\mathscr B}(L^2( \Omega'), L^2( \Omega')) }\leqslant 
| \Im k^2 |^{-1} 
\label{cappone}
\end{equation}
 we have
\begin{equation}
\begin{aligned}
|I_1| & \leqslant  \ve^{\al -2} | \Im k^2 |^{-1}\| g \|_{L^2 }
\lf\| b\lf( \frac{ \cdot}{\ve}, \cdot \ri)\ri\|_{L^{\infty}( \Omega') } 
\lf\| \frac{d}{dt} ( \overline{H}_\ve - k^2 )^{-1} f \ri\|_{L^{2}} \\
& \leqslant  c \ve^{\al -5/2} | \Im k^2 |^{-1}\| g \|_{L^2 }
\| f \|_{L^2}\,,
\end{aligned}
\label{i1}
\end{equation}
where in the second line of \eqref{i1} we have used the fact that there exists $\ve_0$ such that for $0 \leqslant \ve < \ve_0$ we have
$\| b( \frac{ \cdot}{\ve}, \cdot )\|_{L^{\infty}( \Omega')}
\leqslant c < +\infty$, $\ga $ being bounded.

\n
Let us discuss $I_2$; first we notice that 
\begin{equation}\label{fagiano}
\begin{aligned}
&\frac{1}{\ve^2} \lf( V\lf( \frac{t}{\ve}, s\ri) -
\overline{V}\lf(\frac{t}{\ve}\ri) \ri)  =\\
&=\ve^{\al -3} \lf(
\frac{\gamma(t/\ve)^2( 2s\gamma(t/\ve) + \ve^{\al-1} s^2 \gamma(t/\ve)^2 )  }{4(1+\ve^{\al-1}s\gamma(t/\ve))^2}
+\frac{ s\gamma''(t/\ve)}{2(1+\ve^{\al-1}s\gamma(t/\ve))^3}
-\frac{5}{4}\frac{\ve^{\al -1} s^2\gamma'(t/\ve)^2}{(1+\ve^{\al-1}s\gamma(t/\ve))^4}
\ri)
\end{aligned}
\end{equation}
Using the Cauchy-Schwarz inequality, \eqref{fagiano}, \eqref{cappone} and \eqref{tacchino2} we have
\begin{equation}
\begin{aligned}
|I_2| &\leqslant
 | \Im k^2 |^{-1} \| g \|_{L^2 }
\lf\|\frac{1}{\ve^2} \lf( V\lf( \frac{\cdot}{\ve}, \cdot \ri) - \overline{V}\lf(\frac{\cdot}{\ve}\ri) \ri)   
(\Hboh_{\ve} - k^2 - \la_{\ve,m} )^{-1}f \otimes \phi_{m} \ri\|_{L^2( \Omega') }  \\
&\leqslant  | \Im k^2 |^{-1}\| g \|_{L^2 }
\lf\| \frac{1}{\ve^2} \lf( V\lf( \frac{\cdot}{\ve}, \cdot\ri) - \overline{V}\lf(\frac{\cdot}{\ve}\ri) \ri)   \ri\|_{L^2( \Omega') }
\lf\|(\Hboh_{\ve} - k^2 - \la_{\ve,m} )^{-1}f \otimes \phi_{m} \ri\|_{L^{\infty}(\Omega') } \\
& \leqslant c | \Im k^2 |^{-1} \ve^{\al-5/2}\| g \|_{L^2 } \|f\|_{L^2 }
\end{aligned}
\label{i2}
\end{equation}
Estimate \eqref{mulo} follows from \eqref{i1} and \eqref{i2}.
\end{proof}

\n
Now we can prove theorem \ref{mainth}.\\

\n{\bf Proof of theorem \ref{mainth}}.  The proof immediately follows from lemma \ref{lemma1}  and lemma 
\ref{lemma3}. Lemma \ref{lemma3} 
states that  $R_{n,m}^{\ve}(k^2) $ and $  \Rboh_{n,m} (k^2)$ have
the same limit, furthermore
we have $  \Rboh_{n,m} (k^2)= \de_{n,m} ( \overline{H}_{\ve} - k^2 )^{-1}$. Since
$\ga$ has compact support, we can apply lemma \ref{lemma1} to prove the 
uniform convergence of $( \overline{H}_{\ve} - k^2 )^{-1}$ to $( \overline{H}_{\ve}^r - k^2 )^{-1}$
respectively to $( \overline{H}_{\ve}^D - k^2 )^{-1}$ depending on the
presence or not
of a zero energy resonance, and theorem
\ref{mainth} is proved.
\newline  \vspace{.5cm} \noindent\hspace{1cm} \hfill $\square$ \newline

\section{\label{sec3}Properties of the Hamiltonian ${\overline H}^r$ }

\n In this section we characterize the Hamiltonian $\oH^r$. We analyze the
spectrum, we give the explicit expression for the integral kernel of the
propagator and evaluate the scattering matrix.

\n
\begin{proposition}
The operator ${\overline H}^r$ has no point spectrum and no singular continuous spectrum. The
continuous spectrum is $[0,\infty) $ and  there is a zero energy resonance.
\end{proposition}
\begin{proof}
The resolvent \eqref{resHc1c2} has no poles and therefore ${\overline H}^r$ has
no point spectrum. The essential spectrum is $[0,\infty) $ since $\overline{R}^r(k^2)$ is a compact
perturbation of the free resolvent; there is no singular continuous spectrum by theorem XIII.20 in 
\cite{RSIV} and then the continuous spectrum is $[0,\infty) $.

\n
Take two real numbers $a,b$ such that $a(c_1-c_2 ) = b(c_1+c_2)$ and consider $\psi^r$ given by:
\begin{equation*}
\psi^r(t) =
\lf\{
\begin{aligned}
a & \quad t\leqslant 0 \\
b & \quad t> 0
\end{aligned}
\ri.
\end{equation*}
It is straightforward to check that $\psi^r \in L^{\infty} $ and that it is a distributional solution
of ${\overline H}^r \psi^r =0$. In fact take
\begin{equation*}
\begin{aligned}
{\mathscr E} = \big\{ \eta \in {\mathscr D}( {\overline H}^r ) \,\,
s.t. \,\, \eta \in C^{\infty}_0\big(( -\infty, 0]\big) 
\cap C^{\infty}_0\big([0,\infty)\big)\,,\,
&(c_1 + c_2 ) \eta(0^+ ) = (c_1 - c_2 ) \eta(0^- )
\, ,\\ 
&(c_1 - c_2 ) \eta' (0^+ ) = (c_1 + c_2 ) \eta' (0^- )
\big\}\,.
\end{aligned}
\end{equation*}
The set ${\mathscr E}$ is a core for ${\overline H}^r$ and integrating by parts we have:
\begin{equation}
( \eta , {\overline H}^r \psi^r )-
( {\overline H}^r \eta ,  \psi^r )=
a \eta'(0^{-} ) - b \eta( 0^+ ) = 0\,.
\label{weaksol}
\end{equation}
for any $\eta \in {\mathscr E}$.
\end{proof}

\n  The
integral kernel of the one parameter unitary group $e^{-i t \overline{H}^r}$ can be derived 
by using the results of \cite{ABD} and we obtain:
\begin{equation*}
e^{-i t \overline{H}^r } (x,y) = e^{-i t \overline{H}_0 } (x-y) -
 \lf[
\frac{c_2^2}{c_1^2 + c_2^2 } + \frac{ c_1c_2}{c_1^2 + c_2^2 } \sgn x +\frac{ c_1c_2}{c_1^2 + c_2^2 } \sgn y
-\frac{c_2^2}{c_1^2 + c_2^2 } \sgn xy 
\ri]
e^{-i t \overline{H}_0 } (|x|+|y|)
\end{equation*}
where $e^{-i t \overline{H}_0 } (x) = (4 \pi i t )^{-1/2} e^{ -i
  \frac{|x|^2}{4t} } $ is the well know propagator of the free
Schr\"odinger equation. 

\n
It is possible to compute the generalized eigenfunctions: let $p$ a positive number and
let us define two family of functions.
\begin{equation}
\psi_p^{+}(x) = \lf\{
\begin{aligned}
e^{i p x }+ \frac{2c_1 c_2 }{c_1^2 + c_2^2 } e^{-i p x} & \quad x<0 \\
\frac{c_1^2 - c_2^2}{c_1^2 + c_2^2}e^{i p x } & \quad x>0
\end{aligned}
\ri.
\label{psim}
\end{equation}
\begin{equation}
\psi_p^{-}(x) = \lf\{
\begin{aligned}
\frac{c_1^2 - c_2^2}{c_1^2 + c_2^2}e^{-i p x } & \quad x< 0 \\
e^{-i p x }-  \frac{2 c_1 c_2 }{c_1^2 + c_2^2 } e^{i p x} & \quad x>0 
\end{aligned}
\ri.
\label{psip}
\end{equation}
It is straightforward to check that $\psi_p^{+}$ and $\psi_p^{-}$ are
linearly independent and satisfy
\begin{equation*}
\overline{H}^r \psi_p^{\pm} = p^2 \psi_p^{\pm}
\end{equation*}
in a weak sense, as defined in \eqref{weaksol}. It is immediate to compute the reflection and
transmission coefficients ${\mathcal R}_{\pm}$ and ${\mathcal T}_{\pm}$
from \eqref{psip} and \eqref{psim} and we have:
\begin{equation*}
{\mathcal T}_{\pm} = \frac{c_1^2 - c_2^2}{c_1^2 + c_2^2} \qquad
{\mathcal R}_{\pm} = \pm  \frac{2 c_1 c_2 }{c_1^2 + c_2^2 }
\end{equation*}
Notice that ${\mathcal R}_{\pm}$ and ${\mathcal T}_{\pm}$ do not depend on the energy parameter $p$.
The scattering matrix ${\mathcal S}$ is given by:
\begin{equation*}
{\mathcal S} = \lf[
\begin{aligned}
\frac{c_1^2 - c_2^2}{c_1^2 + c_2^2} &\quad - \frac{2 c_1 c_2 }{c_1^2 + c_2^2 } \\
\frac{2 c_1 c_2 }{c_1^2 + c_2^2 }  & \qquad \frac{c_1^2 - c_2^2}{c_1^2 + c_2^2}
\end{aligned}
\ri]
\end{equation*}

\section{\label{sec4}Examples}

\n
In this section we shall present two simple examples  
 and we
shall make some remarks 
about the dependence of the limit operator on the initial curve $\Ga$.

\n Let us discuss some properties of symmetric potentials.  
Assume that $\oV(t)$ satisfies the hypothesis of proposition \ref{prop2}. Assume
moreover that it is such that
the Hamiltonian $\oH$ defined in \eqref{light} has a zero energy
resonance and that $\oV(t)=\oV(-t)$. Let us indicate with $\psi^r(t)$
the resonance of $\oH$. Since $\oV(t)$ is symmetric the function
$\psi^r(t)$ has a definite parity. Given $\psi^r(t)$, the function
$\varphi_0(t)$ solution of equation \eqref{eqphi0}, satisfies
$\varphi_0(t)=-u(t)\psi^r(t)$ a.e. (see Lemma 2.2. in
\cite{BGW}). Since $u(t)$ is symmetric, $\varphi_0(t)$ has the same
parity as $\psi^r(t)$. Then a simple calculation shows that
 only two boundary conditions
  for the functions in the domain of the limit operator $\oH^r$ are possible
\begin{equation*}
\begin{aligned}
&f(0^-)=f(0^+)\,,\quad f'(0^-)=f'(0^+)\,,\qquad
&\psi^r(t)\;\textrm{ even}\\
&f(0^-)=-f(0^+)\,,\quad f'(0^-)=-f'(0^+)\,,\qquad &\psi^r(t)\;\textrm{
  odd}\,.
\end{aligned}
\end{equation*}
Let us notice that if $\psi^r$ is even the  limit operator $\oH^r$
is the free Laplacian.
 
\n Since $\overline{V}=-\ga^2/4$ it is clear that 
the potential does not determine the curvature uniquely. Then we
expect that different curves give the same limit operator. 

\n In the following examples we will consider curves for which the curvature is piecewise constant. Before discussing
 the examples, let us show that, by relaxing the assumption on $\alpha$, it is possible to take into account curves with piecewise constant curvature.

\n Let us consider the curvature
\begin{equation}\label{house}
\gamma(t)=\left\{
\begin{aligned}
&0\quad&&t\notin[a,b]\\
&c_i&&t\in[x_{i-1},x_{i}]\quad i=1,\dots,n
\end{aligned}\right.
\end{equation}
\n where $c_i\in\erre$, and we fix $x_0=a$ and $x_n=b$. We assume that the constants $c_i$ are chosen in such a way that the corresponding curve $\Gamma$ has not self-intersections. The curve corresponding to the curvature \eqref{house} is straight outside $[a,b]$ and in every  interval $[x_{i-1},x_i]$ it is the  arc of a circumference with radius $|c_i|$. Consider $\beta>0$ and $0<\ve<\ve_0$ such that $\ve_0^\beta<1/2\min_{i\neq j}|x_i-x_j|$. Let us define the smoothed version of $\gamma(t)$:
\begin{equation*}
\gsmt_\ve(t)=\left\{
\begin{aligned}
&0\quad&&t\notin[a-\ve^\beta,b+\ve^\beta]\\
&c_i&&t\in[x_{i-1}+\ve^\beta,x_{i}-\ve^\beta]\quad i=1,\dots,n\\
&p_{\ve,i}(t)&&t\in[x_i-\ve^\beta,x_i+\ve^\beta]\quad i=0,1,\dots,n
\end{aligned}\right.
\end{equation*}
\n where the functions $p_{\ve,i}(t)$ are piecewise $C^2([x_i-\ve^\beta,x_i+\ve^\beta])$ and 
\begin{equation*}p_{\ve,i}(x_i-\ve^\beta)=c_i\,,\quad p_{\ve,i}(x_i+\ve^\beta)=c_{i+1}\,,\quad p'_{\ve,i}(x_i\pm\ve^\beta)=0\,;\qquad n=0,1,\dots,n\,,\end{equation*}
\n here $c_0=c_{n+1}=0$. The curvature $\gsmt_\ve(t)$ is piecewise $C^2$, and $\gsmt'_\ve$ and $\gsmt''_\ve$ are bounded for every $\ve>0$ moreover the curve $\Gsmt_\ve$, corresponding to the curvature $\gsmt_\ve$, has no self-intersections. The functions $p_{\ve,i}(t)$ can be chosen, e.g., in the following way
\begin{equation*}
p_{\ve,i}(t)=\left\{\begin{aligned}
&\frac{c_{i+1}-c_i}{2\ve^{2\beta}}(t-x_i)^2+\frac{c_{i+1}-c_i}{\ve^{\beta}}(t-x_i)+
\frac{c_{i+1}+c_i}{2}\quad&t\in[x_i-\ve^\beta,x_i]\\
&-\frac{c_{i+1}-c_i}{2\ve^{2\beta}}(t-x_i)^2+\frac{c_{i+1}-c_i}{\ve^{\beta}}(t-x_i)+
\frac{c_{i+1}+c_i}{2}\quad&t\in[x_i,x_i+\ve^\beta]
\end{aligned}\right.\quad i=0,1,\dots,n\,.
\end{equation*}

\n Let us indicate with $\Osmt_\ve$ the strip of width $2\ve^\alpha d$ defined by ``fattening'' the smoothed curve and by the rescaling $\gsmt_\ve(t)\to\ve^{-1}\gsmt_\ve(t/\ve)$, as it was done in the definition of $\Omega_\ve$, notice that $\Osmt_\ve$ depends on $\ve$ also because of the explicit dependence of $\gsmt_\ve(t)$  on $\ve$. For every $\ve>0$ the operator $-\Delta_{\Osmt_\ve}^D$ is unitarily equivalent to $\Hsmt_\ve$ defined as the closure of the essentially self-adjoint operator $\Hsmt_{0\ve}$
\begin{equation*}
\Hsmt_{0\ve}=-
\pd{}{t}\frac{1}{(1+\ve^{\al-1} s\gsmt_\ve(t/\ve))^2}\pd{}{t}-
 \frac{1}{\ve^{2\al}} \pd{^2}{s^2}+\frac{1}{\ve^2} \Vsmt_\ve(t,s)\,,
\end{equation*}
\n with
\begin{equation*}
\Vsmt_\ve(t,s)=-\frac{\gsmt_\ve(t/\ve)^2}{4(1+\ve^{\al-1}s\gsmt_\ve(t/\ve))^2}
+\frac{\ve^{\al-1} s\gsmt_\ve''(t/\ve)}{2(1+\ve^{\al-1}s\gsmt_\ve(t/\ve))^3}
-\frac{5}{4}\frac{\ve^{2\al -2} s^2\gsmt_\ve'(t/\ve)^2}{(1+\ve^{\al-1}s\gsmt_\ve(t/\ve))^4}
\end{equation*}
and
\begin{equation*}
{\mathscr D}(\Hsmt_{0\ve})=\{
\psi\in L^2(\Omega') \,\, s.t. \,\, \psi\in
C^\infty(\Omega') \,,\,\psi(t,d)=\psi(t,-d)=0 \, , \,
\Hsmt_{0\ve} \psi \in L^2(\Omega') \}\,.
\end{equation*}
\n Let us prove that the result stated in lemma \ref{lemma3} holds if $H_\ve$ is replaced by $\Hsmt_{\ve}$ and $\oV=-\gamma^2/4$ where   $\gamma$  is the piecewise constant function given in  \eqref{house}. Let us define 
\begin{equation*}\Rsmt_{n,m}^{\ve}(k^2, t,t') =  \int_{-d}^{d}ds\,ds' \,
\phi_{n}(s)  ( \Hsmt_{\ve} - k^2 - \la_{\ve,m} )^{-1}(t,s,t',s')
\phi_{m} (s')\end{equation*}
\n and put   $\oV=-\gamma^2/4$ with $\gamma$ defined in \eqref{house}. As it was done in the proof of lemma \ref{lemma3} we need to estimate  
\begin{equation*}
\Ismt_1 =
\lf( g \otimes \phi_{n},
( \Hsmt_{\ve} - k^2 - \la_{\ve,m} )^{-1}
\ve^{\al -2} \bsmt_\ve\lf( \frac{\cdot}{\ve}, \cdot\ri) \pd{}{t}
(\Hboh_{\ve} - k^2 - \la_{\ve,m} )^{-1}f \otimes \phi_{m}\ri)
\end{equation*}
and
\begin{equation*}
\Ismt_2 =
\lf( g \otimes \phi_{n},
( \Hsmt_{\ve} - k^2 - \la_{\ve,m} )^{-1}
\frac{1}{\ve^2} \lf( \Vsmt_\ve\lf( \frac{\cdot}{\ve}, \cdot\ri) -
\overline{V}\lf(\frac{\cdot}{\ve}\ri) \ri) 
(\Hboh_{\ve} - k^2 - \la_{\ve,m} )^{-1}f \otimes \phi_{m}\ri)
\end{equation*}
with $\bsmt_\ve(t,s) = -2 s \gsmt_\ve'(t) / (1 + \ve^{\al -1} s \gsmt_\ve(t) )^3 $. The following estimate for $\Ismt_1$ holds 
\begin{equation*}
|\Ismt_1|
 \leqslant  c \ve^{\al -5/2} | \Im k^2 |^{-1}\| g \|_{L^2 }
\| f \|_{L^2}\lf\| \bsmt_\ve\lf( \frac{ \cdot}{\ve}, \cdot \ri)\ri\|_{L^{\infty}( \Omega') } \,.
\end{equation*}
\n Since $\|\gsmt_\ve'\|_{L^\infty(\Omega ')}\leqslant c\ve^{-\beta}$
\begin{equation*}
|\Ismt_1|
 \leqslant  c \ve^{\al -5/2-\beta} | \Im k^2 |^{-1}\| g \|_{L^2 }\,.
\end{equation*}
\n Let us discuss the term $\Ismt_2$. The following estimate
\begin{equation*}
|\Ismt_2|\leqslant c
 | \Im k^2 |^{-1} \| g \|_{L^2 }
\lf\| \frac{1}{\ve^2} \lf( \Vsmt_\ve\lf( \frac{\cdot}{\ve}, \cdot\ri) - \overline{V}\lf(\frac{\cdot}{\ve}\ri) \ri)   \ri\|_{L^2( \Omega') }\|f\|_{L^2 }\,,
\end{equation*}
\n where
\begin{equation*}
\begin{aligned}
&\frac{1}{\ve^2} \lf( \Vsmt_\ve\lf( \frac{t}{\ve}, s\ri) -
\overline{V}\lf(\frac{t}{\ve}\ri) \ri)  =\\
=&\ve^{\al -3} \lf(
\frac{\gamma(t/\ve)^2( 2s\gsmt_\ve(t/\ve) + \ve^{\al-1} s^2 \gsmt_\ve(t/\ve)^2 )  }{4(1+\ve^{\al-1}s\gsmt_\ve (t/\ve))^2}
+\frac{ s\gsmt_\ve''(t/\ve)}{2(1+\ve^{\al-1}s\gsmt_\ve(t/\ve))^3}
-\frac{5}{4}\frac{\ve^{\al -1} s^2\gsmt_\ve'(t/\ve)^2}{(1+\ve^{\al-1}s\gsmt_\ve(t/\ve))^4}
\ri)+\\
&+\ve^{-2}\lf(\frac{\gamma(t/\ve)^2-\gsmt_\ve(t/\ve)^2}{4(1+\ve^{\al-1}s\gsmt_\ve (t/\ve))^2}\ri)\,,
\end{aligned}
\end{equation*}
\n follows directly from formula \eqref{i2} and from
\begin{equation*}\lf\|(\Hboh_{\ve} - k^2 - \la_{\ve,m} )^{-1}f \otimes \phi_{m} \ri\|_{L^{\infty}(\Omega') } \leqslant c\|f\|_{L^2}\,.\end{equation*}

\n A direct calculation shows that
\begin{equation*}
\lf\| \frac{1}{\ve^2} \lf( \Vsmt_\ve\lf( \frac{\cdot}{\ve}, \cdot\ri) - \overline{V}\lf(\frac{\cdot}{\ve}\ri) \ri)   \ri\|_{L^2( \Omega')} \leqslant 
c \ve^{\frac{\beta-3}{2}}+c' \ve^{(2\alpha-3\beta-5)/2}
\end{equation*}
\n where we used $\|\gsmt_\ve'\|_{L^\infty}\leqslant c\ve^{-\beta}$ and $\|\gsmt_\ve''\|_{L^\infty}\leqslant c\ve^{-2\beta}$. Then for $\alpha>5/2+3\beta/2$ and $\beta>3$ the following limit holds
\begin{equation*}
\ulim_{\ve \rightarrow 0}  \big(\Rsmt_{n,m}^{\ve}(k^2) -\Rboh_{n,m}
(k^2)\big)  = 0
\qquad k^2\in\CO\backslash\RE,\;\Im k>0\,.
\end{equation*} 
\n The result of lemma \ref{lemma1} holds for the piecewise constant potential
 $\oV=-\gamma^2/4$ with $\gamma$ defined in \eqref{house}. 
Then in  the following examples we can consider curves with piecewise constant
 curvature such that the one dimensional Hamiltonian $\oH=\oH_0-\gamma^2/4$ has a zero energy resonance,
 such examples should be read by tacking into 
account our comment on the smoothing of curves with piecewise constant curvature.

\n Let us
discuss a simple example of a one parameter family of curves generating the
same symmetric potential. 

\begin{example} The single square well, curves with  fixed curvature
  radius.

\n
Let us consider the potential $\overline V(t)$  defined in the following way:
\begin{equation}
{\overline V}(t)  = \lf\{
\begin{aligned}
0 & \quad t \leqslant 0 \\
-a^2 &  \quad 0<t< b \\
0 &\quad  t\geqslant b
\end{aligned}
\ri.
\label{res1}
\end{equation}
where $a$ and $b$ are positive real numbers.
It is straightforward to prove that $\overline H$ has a zero energy
resonance if and only if  $ab = n \pi $, $n=1,2, \ldots$.  In 
particular take $ab=\pi$ and let us consider the one parameter 
family of curves, $\Gamma_x$, with curvature $\ga_{x}$  defined by:
\begin{equation}
\ga_x(t)  = \lf\{
\begin{aligned}
0 & \quad t \leqslant 0 \\
2a & \quad 0 \leqslant t < x \\
-2a& \quad  x\leqslant t< b \\
0 & \quad t\geqslant b
\end{aligned}
\ri.
\label{resga1}
\end{equation}
\n with $b/4 <x <3b/4$. The restriction on the parameter $x$ avoids
self-intersections.  All the functions $\ga_{x}$ give the potential
\eqref{res1} and we have $\theta(x) = \int_{\erre} \ga_{x}(t) \,dt = 
2a(2x-b) $. Then it straightforward to notice that 
$\theta$ can assume any value between -$\pi$  and $\pi$.
\end{example}
\n 
This example shows
that the angle $\theta$ is not sufficient to characterize the limit Hamiltonian since there are 
infinitely many different
curves with different $\theta$ which have the same limit Hamiltonian $\overline{H}^r$.

\n
The previous example suggests that there is an even greater freedom in constructing
different curves which gives the same limit operator: in facts
for any integer $k>0$ and any partition ${\mathscr P}$ of the interval $(0,b)$ 
into $k$ sub intervals, we can construct a piecewise constant
curvature $\ga_{\mathscr P}$; if
the corresponding curve $\Ga_{\mathscr P}$ has
 no self-intersections, then it satisfies the hypothesis of our theorem. 
All the  $\Ga_{\mathscr P}$ yield the same limit Hamiltonian $\overline{H}^r$
since they have the same resonant potential $\overline{V}$ and generically
these curves will have different $\theta$.

\n
In the previous example it was crucial that $\Ga$ had a turning point where $\ga$ changes sign, otherwise
the curve $\Ga$ would have self-intersections. In the following
example we shall consider  curves such that
$\ga$ has constant sign. Notice that
for this class of curves, the potential $\overline{V}$ uniquely determines
the curvature. We shall see that also with this restriction the angle $\theta$ 
is not sufficient to characterize the limit Hamiltonian $\overline{H}^r$.

\begin{example}The triple (asymmetric) square
  well, curves with fixed signum of the curvature.

\n
Let us consider the following potential:
\begin{equation}
{\overline V}(t)  = \lf\{
\begin{aligned}
0 & \quad t \leqslant 0 \\
-a_1^2 &  \quad -b_1<t< 0 \\
-a_2^2 &\quad 0\leqslant t< b_2 \\
-a_3^2 & \quad b_2 \leqslant t< b_2+b_3 \\
0 & \quad t\geqslant b_2+b_3
\end{aligned}
\ri.
\label{res2}
\end{equation}
where $a_1 , a_2, a_3 , b_1 , b_2 , b_3$ are real positive numbers. In
this example we consider only curves with fixed signum of the
curvature, then we assume that the curvature associated to the potential
\eqref{res2} is 
\begin{equation*}
\gamma(t)  = \lf\{
\begin{aligned}
0 & \quad t \leqslant 0 \\
2a_1 &  \quad -b_1<t< 0 \\
2a_2 &\quad 0\leqslant t< b_2 \\
2a_3 & \quad b_2 \leqslant t< b_2+b_3 \\
0 & \quad t\geqslant b_2+b_3
\end{aligned}
\ri.
\end{equation*}
It is straightforward but tedious to prove that $\overline{H}$ has a zero energy resonance if and only
if the following equation is satisfied:
\begin{equation}
\begin{aligned}
& a_1 a_3 \sin(a_1 b_1 ) \sin(a_2 b_2 ) \sin(a_3 b_3 ) - a_2 a_3  \cos(a_1 b_1 ) \cos(a_2 b_2 ) \sin(a_3 b_3 ) \\
& - a_2^2 \cos(a_1 b_1 ) \sin(a_2 b_2 ) \cos(a_3 b_3 ) - a_1 a_2  \sin(a_1 b_1 ) \cos(a_2 b_2 ) \cos(a_3 b_3 )=0\,.
\end{aligned}
\label{caciotta}
\end{equation}
Since the curvature has definite positive signum, the assumption that
the curve is not self-intersecting is equivalent to the condition
\begin{equation}
\theta = 2( a_1 b_1 + a_2 b_2 +a_3 b_3) < \pi\,.
\label{tetares}
\end{equation}
Then we look for solutions of the equation \eqref{caciotta} satisfying
the condition \eqref{tetares}. As a consequence we have that $0<\cos
(a_i b_i)<1$, 
$i=1,2,3$, 
in such a case  equation \eqref{caciotta} is equivalent to: 
 \begin{equation}
a_1 a_3 \tan(a_1 b_1 ) \tan(a_2 b_2 ) \tan(a_3 b_3 ) - a_2 a_3 \tan(a_3 b_3 )
- a_2^2  \tan(a_2 b_2 )  - a_1 a_2  \tan(a_1 b_1 ) =0\,.
\label{caciotta2}
\end{equation}
It is straightforward to provide infinitely many solutions of
\eqref{caciotta2}. 
Fix $a_1 b_1=\beta_1$, $  a_2 b_2=\beta_2$ and $ a_3 b_3=\beta_3$ such
that, $\beta_1$, $\beta_2$ and $\beta_3$ satisfy the condition \eqref{tetares}; this can be done fixing
$ b_1$, $  b_2$ and $  b_3$ 
 leaving $a_1 $, $  a_2 $ and $ a_3 $ free. Now equation \eqref{caciotta2}
becomes an equation in $a_1 $, $  a_2 $ and $ a_3 $ since 
 $\tan(\beta_1  )$, $\tan(\beta_2 )$ and $\tan(\beta_3 )$ are fixed positive numbers; for instance
we can solve it with respect to $a_1$ and we obtain:
\begin{equation*}
a_1 = \frac{ a_2^2  \tan(\beta_2 )+ a_2 a_3 \tan(\beta_3 ) }{
 \tan(\beta_1 )( a_3  \tan(\beta_2  ) \tan(\beta_3 )-  a_2 ) }\,.
\end{equation*}
Every $a_2$ and $a_3$ such that $ a_3  \tan(\beta_2 ) \tan(\beta_3  )-  a_2 >0$ provide a solution of 
\eqref{caciotta} with a different potential $\overline{V}$ but the
same angle $\theta$. 
\end{example}

\n
Therefore we have showed that there are infinitely many different
curves with the same angle
$\theta$ which give different resonant potentials \eqref{resga1} and 
therefore different limit Hamiltonian $\overline{H}^r$.

\n
It is an interesting open question to find which quantities of the curve $\Ga$ are sufficient to characterize
the limit Hamiltonian $\overline{H}^r$.\\

\end{document}